\documentclass{raa}
\usepackage{graphicx,times,rotating}
\input{epsf.sty}
\input{psfig.sty}

\begin{document}

\title{Near-infrared Observations of Be/X-ray Binary Pulsar A0535+262}

\author{Sachindra Naik, Blesson Mathew, D. P. K. Banerjee, N. M. Ashok \and Rajeev R. Jaiswal}

\institute{ Astronomy and Astrophysics Division, Physical Research Laboratory, Ahmedabad, India\\
              {\it snaik@prl.res.in}
}

\abstract{We present results obtained from an extensive near-infrared spectroscopic
and photometric observations of the Be/X-ray binary A0535+262/HDE~245770 at
different phases of its $\sim$111 day orbital period. This observation
campaign is a part of the monitoring programme of selective Be/X-ray binary
systems aimed at understanding the X-ray and near-IR properties at different
orbital phases, especially during the perisastron passage of the neutron
star. The  near-IR observations presented here were carried out
using the 1.2 m telescope at Mt. Abu IR observatory. Though the source was
relatively faint for spectroscopic observations with the 1.2 m telescope,
we monitored the source closely during the 2011 February--March giant X-ray
outburst to primarily investigate whether any drastic changes in the near-IR $JHK$
spectra take place at the periastron passage. Changes of such a striking nature were 
expected to be detectable in our spectra.  Photometric observations of
the Be star show a gradual and systematic fading in the $JHK$ light curves
since the onset of the X-ray outburst that could suggest a mild evacuation/truncation
of the circumstellar disc of the Be companion. Near-IR spectroscopy of the
object shows that the $JHK$ spectra are dominated by the emission lines of
hydrogen Brackett and Paschen series and HeI lines at 1.0830 $\mu$m, 1.7002
$\mu$m~ and 2.0585 $\mu$m. The presence of all hydrogen emission lines in
the $JHK$ spectra, along with the absence of any significant change in the
continuum  of the Be companion during X-ray quiescent and X-ray outburst
phases suggest that the near-IR line emitting regions of the disc are not
significantly affected during the X-ray outburst.
}

\authorrunning{Naik et al. }
\titlerunning{Near-infrared Observations of A0535+262}

\maketitle

\section{Introduction}

\begin{figure*}
\centering
\includegraphics[height=5.8in, width=2.3in, angle=-90]{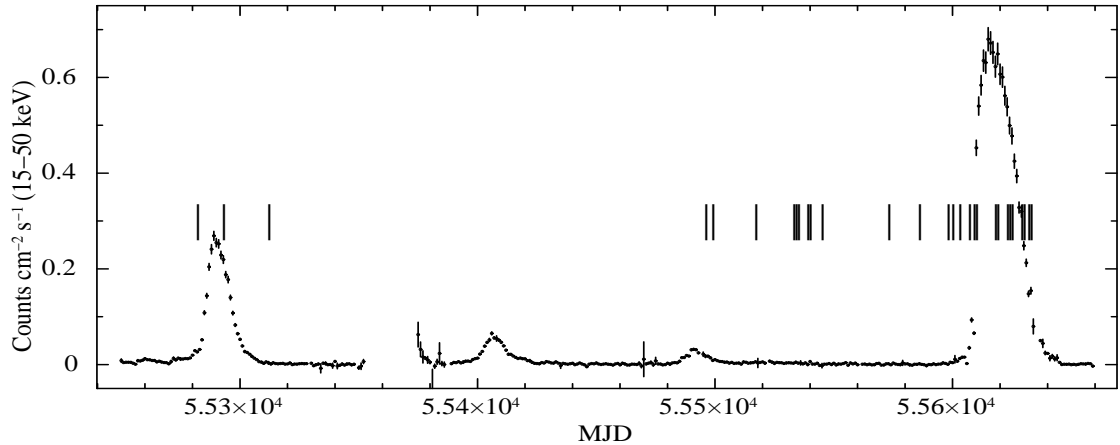}
\caption{The $Swift$/BAT light curve of A0535+262 in 15-50 keV energy band,
from 2010 February 23 (MJD 55250) to 2011 April 08 (MJD 55659). The regular
and periodic X-ray outbursts in the transient Be/X-ray binary pulsar are seen
in the light curve. The epochs of our near-infrared observations are marked by
vertical lines in the figure.}
\label{x-lc}
\end{figure*}

High mass X-ray binary (HMXB) systems are strong X-ray emitters via the
accretion of matter from the OB companion onto the neutron star. These objects
appear as the brightest objects in the X-ray sky. The HMXBs are classified
as Be/X-ray binaries and supergiant X-ray binaries. The Be/X-ray binaries
represent the largest subclass of HMXBs. In Be/X-ray binaries, the optical
companion is a Be star. These Be stars are fast rotating B-type stars
which show spectral lines such as hydrogen lines (Balmer and Paschen series)
in emission (Porter \& Rivinius 2003 and references therein). Apart from
hydrogen lines, the stars occasionally show He, Fe lines in emission (Hanuschik
1996). These objects also show infrared excess i.e an amount of IR radiation 
that is larger than that expected from an absorption-line B star of the same 
spectral type. The origin of the emission lines and infrared excess in Be/X-ray 
binary systems are attributed to the presence of a circumstellar gaseous component 
around the Be star that is commonly accepted to be in the form of equatorial disc. 
The disc is believed to be fed from the material expelled from the rapidly rotating 
Be star (Porter \& Rivinius 2003). The orbit of Be/X-ray binary systems is generally 
wide (with orbital period in the range of tens of days to several hundred days) and 
eccentric (with eccentricity $e$ in the range of 0.1 to 0.9) (Reig 2011). The neutron
star in these Be/X-ray binary systems spends most of the time far away from
the circumstellar disc surrounding the optical companion. Mass transfer takes
place from the Be companion to the neutron star through the circumstellar
disc. Strong X-ray outbursts are normally seen when the neutron star
(pulsar) passes through the circumstellar disc or during the periastron
passage (Okazaki \& Negueruela 2001). The X-ray emission of such systems
can be transiently enhanced by a factor of $\sim$10 and more. Each of the
Be/X-ray binary systems show periodic (Type I) X-ray outbursts that coincide
with the periastron passage, giant (Type II) X-ray outbursts which do not show
any clear orbital modulation, and/or persistent low luminosity X-ray emission
(Negueruela et al. 1998). The Be/X-ray binaries, therefore, attract special
interests in several branches of astrophysics (viz. X-ray, optical, infrared
bands etc.) to study the effect of the neutron star on the circumstellar disc
of the Be star companion.

A0535+262 is a 103 s Be/X-ray binary pulsar discovered by $Ariel~V$ during
a large (Type II) outburst in 1975 (Coe et al. 1975). The binary companion
HDE~245770 is an O9.7-B0 IIIe star in a relatively wide eccentric orbit ($e$
= 0.47) with orbital period of $\sim$111 days and at a distance of $\sim$2 kpc
(Finger et al. 1996; Steele et al. 1998). The pulsar shows regular outbursts
with the orbital periodicity. Occasional giant X-ray outbursts are also observed
when the object becomes even brighter than the Crab (Naik et al. (2008) and 
references therein). The pulsar shows three typical intensity states, such 
as quiescence with flux level of below 10 mCrab, normal outbursts with flux 
level in the range 10 mCrab to 1 Crab, and giant outbursts during which the 
object becomes the brightest X-ray source in the sky with the flux level of 
several Crab (Kendziorra et al. 1994). Extensive photometric and spectroscopic 
work in the ultra-violet, optical and infrared bands show the variable nature 
of the optical companion of the pulsar (Clark et al. 1998a; Haigh, Coe \& 
Fabregat 2004). Infrared spectroscopy of the Be companion HDE~245770 obtained 
over 1992-1995 showed significant variability implying changes in the circumstellar 
disc (Clark et al. 1998b). A decrease in the flux of Paschen series lines, 
strength of H$\alpha$ line and the optical continuum emission were seen between 
1993 December and 1994 September. These changes were attributed to the reduction 
in the emission measure of the Be disc (Clark et al. 1998b).

A striking episode of circumstellar disc loss and subsequent formation of an 
new inner disc in the Be binary system A0535+262/HDE~245770 has been reported 
earlier (Haigh et al. 1999). The Br$\gamma$ emission line which was earlier seen 
in emission had gone into absorption, as detected on 1998 November 10 (Figure~4 
of Haigh et al. 1999). Along with the change in Br$\gamma$ line (from emission 
to absorption), the HeI line at 2.058 $\mu$m was also detected in emission
with significantly reduced intensity. During this particular disc-loss phase,
the H$\alpha$ emission line was also found to be absent. Following the complete
loss of emission, symmetrical emission wings were formed in H$\alpha$ and the
spectra appeared to remain stable for a few months. The disc-loss state in the
Be star HDE~245770, however, has not been seen again. High-dispersion optical
spectroscopic observations of the Be star, during a giant X-ray outburst in
2009 November--December, suggested the presence of the active components in the
Be circumstellar disc that causes the observed significant variability
in the emission line profiles (Moritani et al. 2011). The detection of H$\alpha$
line in emission, during the giant X-ray outburst, suggest that the disc-loss is
not as significant as observed in November 1998.

During the periastron passage of the neutron star in Be binary systems,
the circumstellar disc of the Be companion is expected to be most affected.
As the contribution of the circumstellar disc towards the total infrared
emission from the system is large, the effect of the periastron passage
should be pronounced in the infrared rather than optical bands. Hence our 
motivation for the present IR studies. Our campaign covered the X-ray quiescent 
and outburst phases of the binary system spanned over four orbital cycles. It 
may be noted that such a contemporaneous IR coverage of the giant X-ray burst,  
as presented here, has not been undertaken earlier.

\begin{table*}
 \centering
\caption{Log of the Mt. Abu near-infrared observations of the Be star in A0535+262/HDE~245770 binary system.}
\begin{tabular}{@{}lccccccccc@{}}
\hline
\hline
     &\multicolumn{3}{|c|}{Spectroscopic Observations}   &\multicolumn{6}{|c|}{Photometric Observations}  \\
\hline
 Date of		&\multicolumn{3}{|c|}{Integration time (s)} &\multicolumn{3}{|c|}{Integration time (s)} &\multicolumn{3}{|c|}{Magnitude}\\
 Observation      		& J-band & H-band & K-band	 	& J-band  &H-band  &K-band    & J-band  &H-band  &K-band \\
\hline
2010 Mar. 28 	& 200 & 300 & 300 	&230  &280  &210 	& 7.93$\pm$0.04 	& 7.68$\pm$0.03  	 &7.26$\pm$0.06\\
2010 Apr. 08 	& 300 & 300 & 300	& --- & --- & ---	& --- 			& --- 			& ---\\
2010 Apr. 27 	& --- & --- & ---	&225  &225  &53 	& 8.36$\pm$0.07 	& 7.93$\pm$0.05  	 &7.59$\pm$0.37\\
2010 Oct. 28 	& --- & --- & ---	&250  &250  &250	& 7.69$\pm$0.02 	& 7.63$\pm$0.03  	 &7.42$\pm$0.02\\
2010 Oct. 31 	& 100 & 120 & 120	&150  &150  &250	& 7.62$\pm$0.03 	& 7.50$\pm$0.02  	 &7.39$\pm$0.02\\
2010 Nov. 18 	& 190 & 190 & 190 	& --- & --- & ---       & --- 			& --- 			& --- \\
2010 Dec. 04 	& 120 & 120 & 120 	& --- & --- & ---       & --- 			& --- 			& ---\\
2010 Dec. 05 	& 120 & 120 & 120 	& --- & --- & ---       & --- 			& --- 			& ---\\
2010 Dec. 06 	& 120 & 120 & 120 	& --- & --- & ---       & --- 			& --- 			& ---\\
2010 Dec. 10 	& --- & --- & ---	&150  &150  &250	& 7.87$\pm$0.03 	& 7.65$\pm$0.03  	 &7.44$\pm$0.08\\
2010 Dec. 11 	& 120 & 120 & 120	& --- & --- & ---       & --- 			& --- 			& ---\\
2010 Dec. 16 	& 120 & 120 & 120 	&150  &150  &250	& 7.90$\pm$0.02 	& 7.70$\pm$0.02  	 &7.48$\pm$0.03\\
2011 Jan. 13 	& 120 & 120 & 120 	&150  &150  &250	& 7.81$\pm$0.01 	& 7.62$\pm$0.05  	 &7.49$\pm$0.06\\
2011 Jan. 26 	& 120 & 120 & 120 	&100  &100  &250	& 7.98$\pm$0.06 	& 7.63$\pm$0.02 	 &7.39$\pm$0.07\\
2011 Feb. 07 	& --- & --- & ---	&100  &100  &250	& 7.85$\pm$0.02 	& 7.64$\pm$0.01  	 &7.47$\pm$0.02\\
2011 Feb. 09 	& 180 & 180 & 180 	&100  &100  &250	& 7.90$\pm$0.02 	& 7.73$\pm$0.06  	 &7.46$\pm$0.03\\
2011 Feb. 12 	& --- & --- & ---	&150  &150  &250	& 7.89$\pm$0.01 	& 7.64$\pm$0.03  	 &7.45$\pm$0.02\\
2011 Feb. 16 	& 180 & 180 & 150 	&50   &50   &25		& 7.91$\pm$0.04 	& 7.66$\pm$0.03 	 &7.45$\pm$0.04\\
2011 Feb. 18 	& 180 & 180 & 180 	&150  &150  &250	& 7.94$\pm$0.03 	& 7.76$\pm$0.02  	 &7.47$\pm$0.03\\
2011 Feb. 19 	& 180 & 240 & 240 	&100  &100  &250	& 7.98$\pm$0.04 	& 7.65$\pm$0.04  	 &7.46$\pm$0.02\\
2011 Feb. 27 	& 180 & 120 & 120 	&100  &80   &200	& 7.85$\pm$0.02 	& 7.71$\pm$0.04  	 &7.45$\pm$0.04\\
2011 Feb. 28 	& 180 & 180 & 150 	&100  &100  &250	& 7.83$\pm$0.02 	& 7.69$\pm$0.04  	 &7.51$\pm$0.03\\
2011 Mar. 04 	& 180 & 180 & 180 	&110  &165  &250	& 8.02$\pm$0.03 	& 7.71$\pm$0.02  	 &7.54$\pm$0.01\\
2011 Mar. 05 	& 180 & 180 & 180 	&110  &165  &250	& 7.99$\pm$0.01 	& 7.73$\pm$0.03  	 &7.54$\pm$0.01\\
2011 Mar. 06 	& 180 & 180 & 180 	&110  &165  &105	& 7.95$\pm$0.01 	& 7.78$\pm$0.02  	 &7.48$\pm$0.02\\
2011 Mar. 10 	& --- & 180 & 180 	&110  &165  &105	& 7.98$\pm$0.02 	& 7.77$\pm$0.03  	 &7.52$\pm$0.02\\
2011 Mar. 11 	& 180 & 180 & 180 	&110  &165  &105	& 7.99$\pm$0.01 	& 7.76$\pm$0.01  	 &7.55$\pm$0.05\\
2011 Mar. 13 	& --- & --- & ---	&50   &50   &250	& 7.99$\pm$0.02 	& 7.71$\pm$0.02  	 &7.54$\pm$0.03\\
2011 Mar. 14 	& 180 & 180 & 180 	& --- & --- & ---       & --- 			& --- 			& ---\\
\hline
\hline
\end{tabular}
\end{table*}

\begin{figure}
\centering
\includegraphics[height=5.7in, width=4.3in, angle=-90]{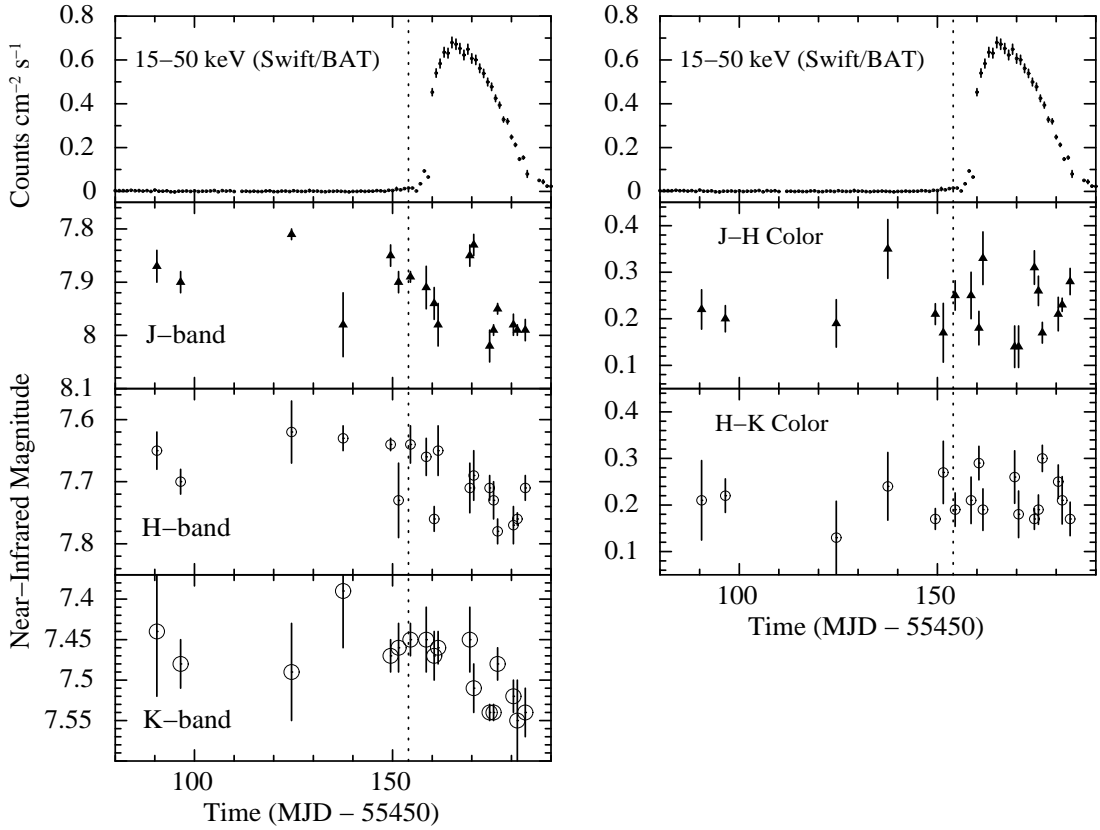}
\caption{The Swift/BAT X-ray light curve (in 15-50 keV energy band; top panels)
and the near-infrared $JHK$ light curves (left panels) of the Be star in 
A0535+262/HDE~245770 binary system, covering the recent X-ray outburst in 2011 
February-March. The second and third panels in right side show the J-H and H-K
colors during the quiescent and outburst phase of the Be/X-ray binary system.
The dotted line indicates the onset of the 2011 February--March X-ray outburst.}
\label{jhk-lc}
\end{figure}

\section{Observations and Data Reduction}

Near-infrared spectroscopic and photometric observations of Be star 
HDE~245770 were carried out by using the 1.2-m telescope of the Mt. 
Abu Infrared Observatory. As in case of Be/X-ray binaries, regular 
X-ray outbursts of the Be binary pulsar A0535+262 are detected each 
time the neutron star undergoes periastron passage using the monitoring 
detectors onboard various X-ray observatories such as MAXI/GSC, RXTE/ASM, 
Swift/BAT, INTEGRAL etc. (Nakajima et al. 2010; Mihara et al. 2010; 
Caballero et al. 2011; Tchernin et al. 2011). Following the detection of 
X-ray outbursts, the Be star in A0535+262/HDE~245770 binary system was 
observed in radio, optical, infra-red bands at different epochs, the 
results of which are reported in literature (Giovannelli et al. 2010; 
Tudose et al. 2010; Mathew et al. 2010; Migliari et al. 2011). We carried 
out photometric and spectroscopic observations of the Be companion
at different orbital phases spanned over three binary periods.

The $Swift$/BAT X-ray light curve of the pulsar in the bianry system 
covering $\sim$4 orbital cycles (from 2010 February 23 to 2011 April 
08), in 15--50 keV energy range, is shown in Figure~\ref{x-lc} with the 
epochs of our near-infrared observations marked by vertical lines - the 
log of the observations is given in Table~1. The signal-to-noise 
ratio (S/N) of the spectral observations were in the range of 20-30, 
25-40 and 15-30 for $J$, $H$ and $K$-bands, respectively. The Mt. Abu 
spectra were obtained at a resolution of $\sim$1000 using a Near-Infrared 
Imager/Spectrometer with a 256$\times$256 HgCdTe Near-Infrared 
Camera Multiobject Spectrograph 3 (NICMOS3) array. Photometric 
observations of the Be star were carried out on several nights 
(Table~1) in photometric sky conditions using the NICMOS3 array 
in the imaging mode. Several frames were obtained in five dithered
positions, typically offset by $\sim$30 arcsec from each other, 
with exposure times ranging from 1--7 s depending on the brightness 
of the object in $JHK$ bands. The sky frames were generated by median 
combining the average of each set of dithered frames and subsequently 
subtracted from the source frames. A nearby field-star~SAO 77466,
observed at similar airmass as the Be/X-ray binary, was used as the 
standard star for photometric observations. Aperture photometry was 
done using the APPHOT task in IRAF.

Spectral calibration was done using the OH sky lines that register with 
the stellar spectra. The spectra of the nearby field star SAO~76920 were 
taken in $JHK$ bands at similar airmass as that of A0535+262/HDE~245770 
on all the observation nights to ensure that the ratioing process (Be 
star spectrum divided by the standard star spectrum) removes the telluric 
lines reliably. The ratioed spectra were then multiplied by a blackbody 
curve corresponding to the standard star's effective temperature to yield 
the final spectra. The detailed reduction of the spectral and photometric 
data, using IRAF tasks, follows a standard procedure that is described in 
Naik et al. (2009, 2010).

\begin{figure*}
\vskip 10.5 cm
\includegraphics{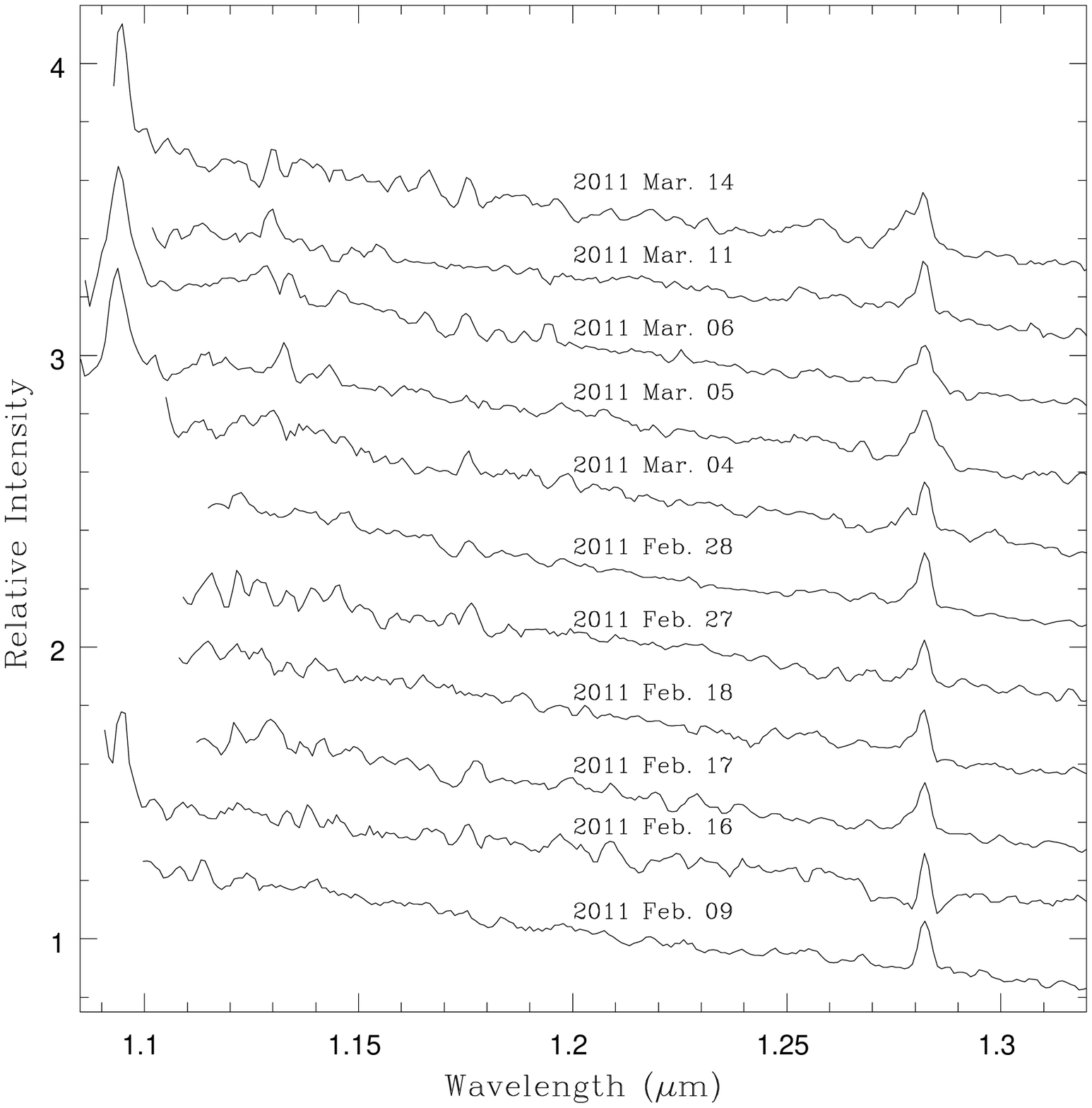}
\includegraphics{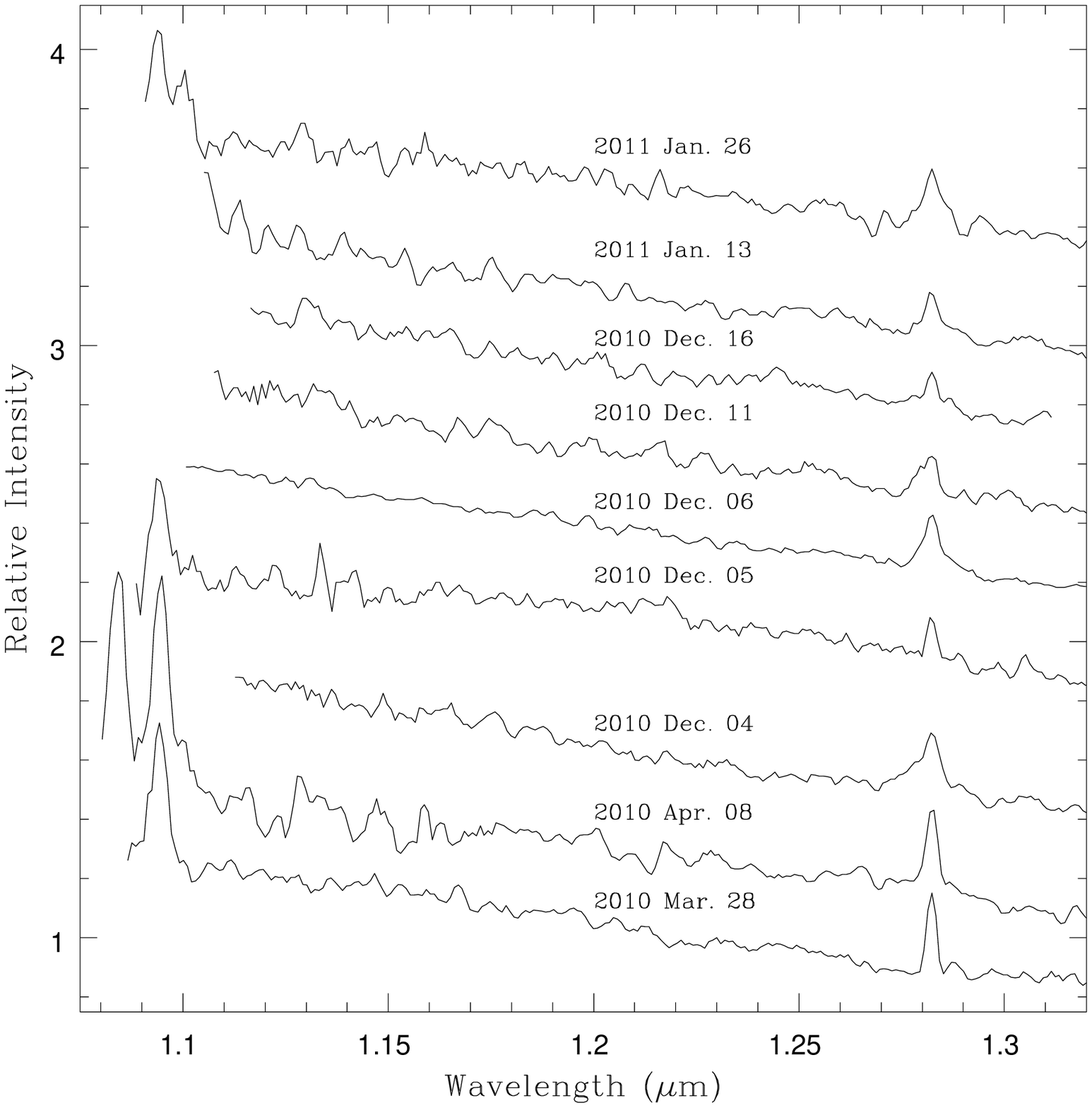}
\caption{The $J$-band spectra of A0535+262/HDE~245770 at different epochs with the
continuum being normalized to unity at 1.22 $\mu$m. }
\label{J-band}
\end{figure*}

\section{Results and Discussion}

\subsection{X-ray and infrared $JHK$ light curves of A0535+262/HDE~245770 
binary system}

The $JHK$ light curves of the optical companion of the transient Be/X-ray
binary pulsar A0535+262, obtained from the present photometric observations
are presented in second, third and fourth panels in left side of
Figure~\ref{jhk-lc}. The $Swift$/BAT X-ray light curve of the pulsar (in 
15-50 keV energy range) covering the 2011 February--March outburst is 
also shown in the top panels (left and right sides) of the figure 
to compare the changes in $JHK$ magnitudes of the companion and corresponding 
changes in the X-ray intensity during the periastron passage of the neutron 
star. During X-ray quiescent, the $JHK$ magnitudes of the companion remain 
almost constant (within errorbars). This represents little changes in the 
infrared emission during the X-ray quiescent viz. when the neutron star is 
far away from the Be companion in the binary orbit. However, a gradual and 
systematic change in the $JHK$ magnitudes of the Be companion is seen since 
the onset of the X-ray outburst (as marked with dotted lines in 
Figure~\ref{jhk-lc}). Simultaneous increase in the X-ray brightness of the 
neutron star corroborates the circumstellar disc evacuation/truncation. 
The $J-H$ and $H-K$ colors, as shown in the second and third panels 
in right side of Figure~\ref{jhk-lc}, do not show any systematic variation 
during the X-ray outburst. The $J-H$ and $H-K$ values remain constant 
(within errors) during the quiescent as well as outburst phase of the binary
orbital period. This suggests that the reduction in the near-infrared flux 
during the 2011 February--March X-ray outburst is approximately same in 
$JHK$ bands.

It is known that the Be circumstellar disc contributes significantly, 
through free-free and bound-free emission,  towards the infrared emission 
from the Be star system. The observed fading in the $JHK$ magnitudes of 
the Be companion in A0535+262/HDE~245770 binary system, therefore, can be 
interpreted as the possible evacuation/truncation of the circumstellar 
disc around the Be star during the periastron passage of the neutron star. 
During the X-ray outburst in 2011 February--March, the decrease in the 
observed $JHK$ photometric magnitudes of the Be star companion is found 
to be $\sim$0.12. The change in magnitude of $\sim$0.12 in $JHK$-bands 
implies a reduction in the source flux by $\sim$12\% that arises possibly 
because of evacuation/truncation of the circumstellar disc during the 
periastron passage of the neutron star. The amount of matter evacuated
from the circumstellar disc of the Be star represents the magnitude of 
the observed X-ray outburst.

\begin{figure*}
\vskip 10.5 cm
\includegraphics{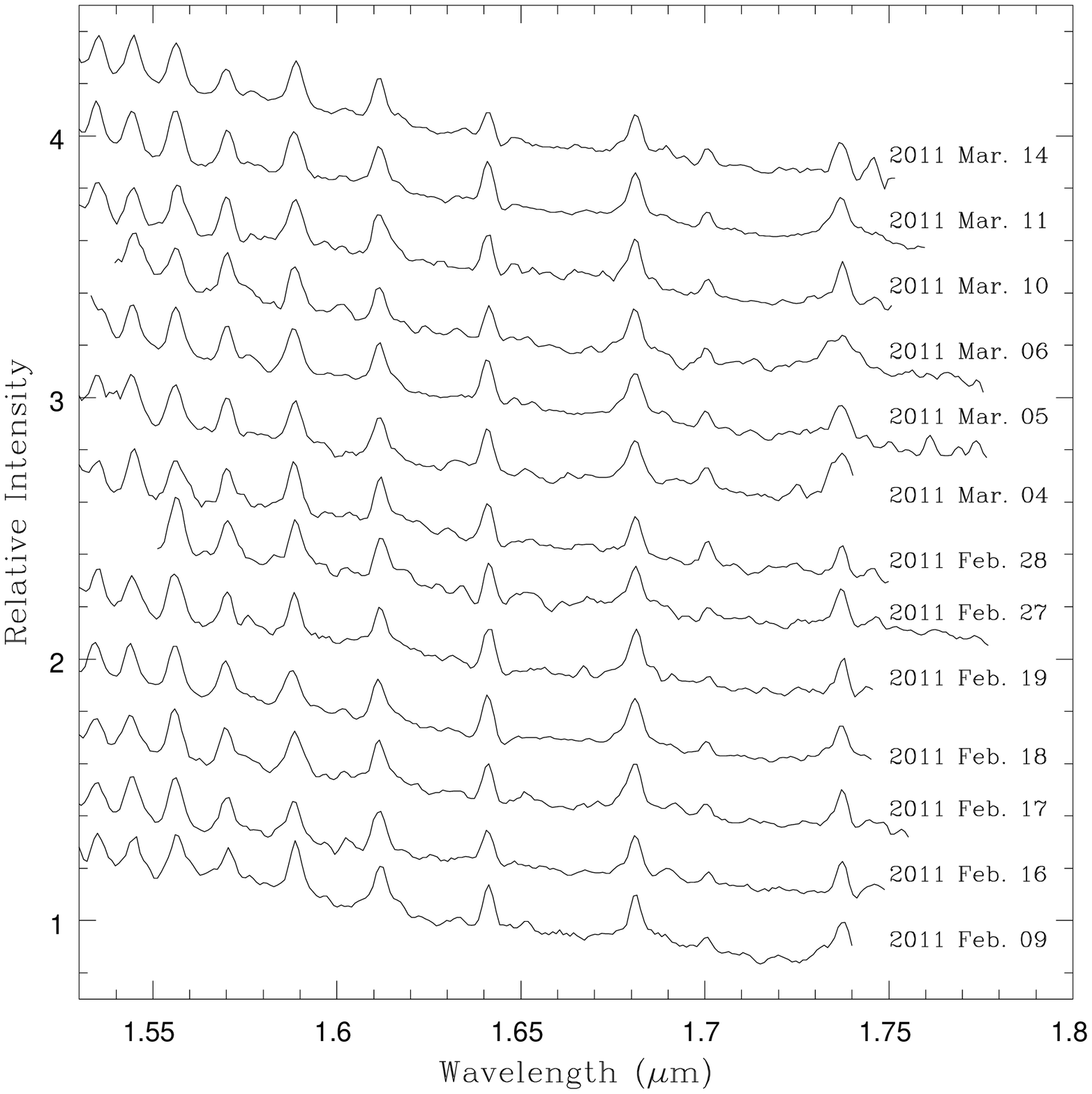}
\includegraphics{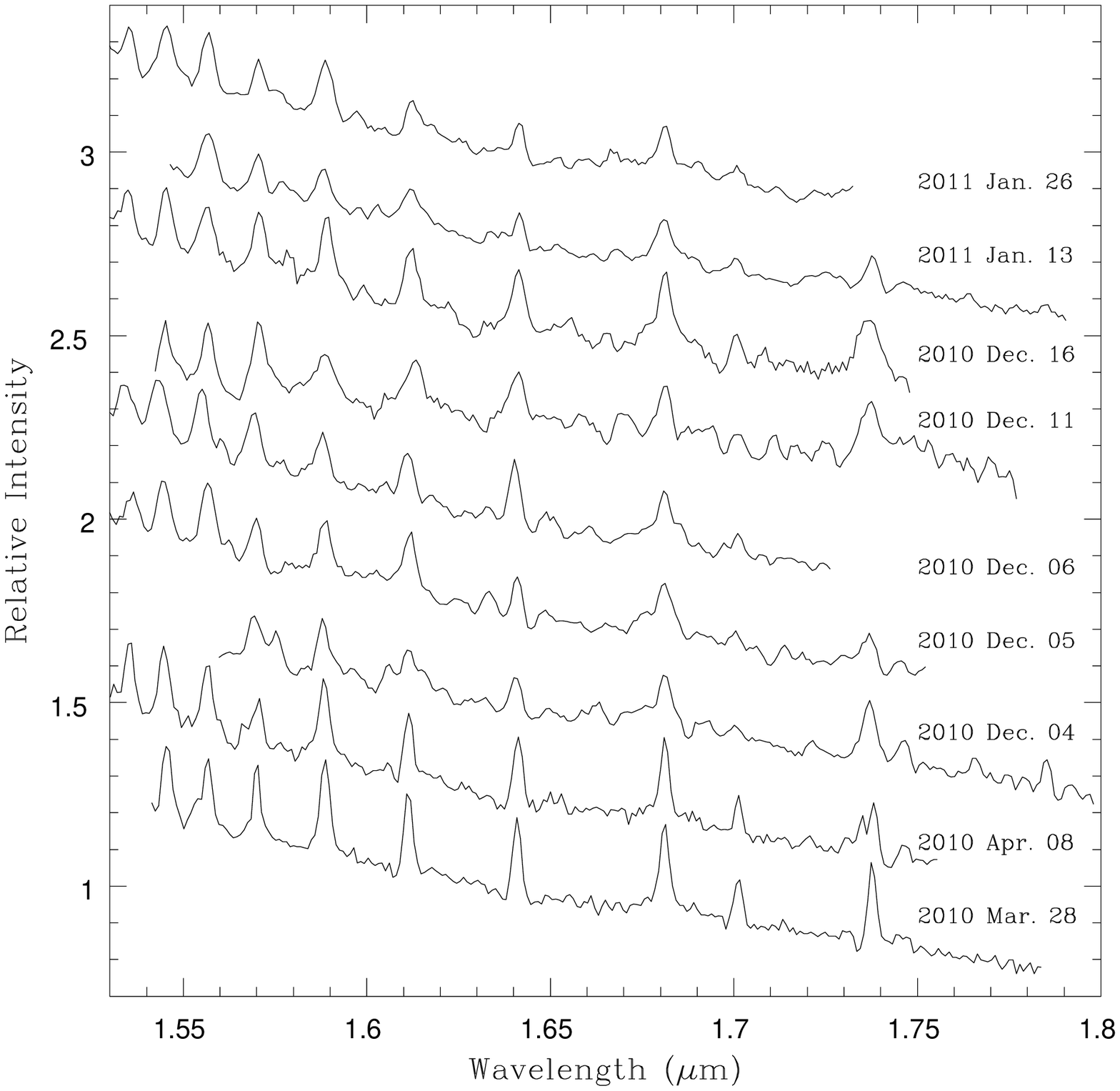}
\caption{The $H$-band spectra of A0535+262/HDE~245770 at different epochs with the
continuum being normalized to unity at 1.63 $\mu$m. }
\label{H-band}
\end{figure*}

\begin{figure*}
\vskip 10.5 cm
\includegraphics{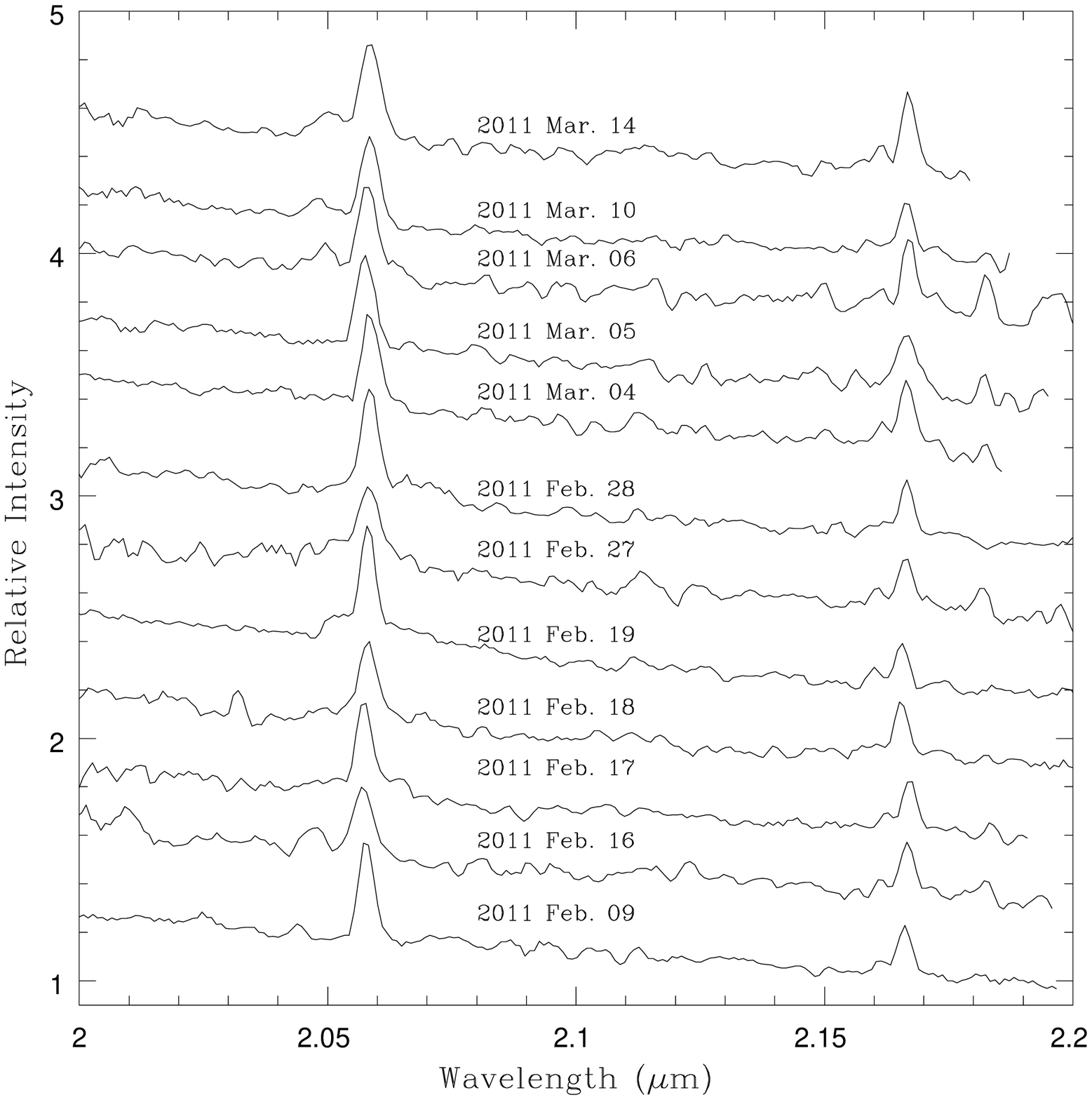}
\includegraphics{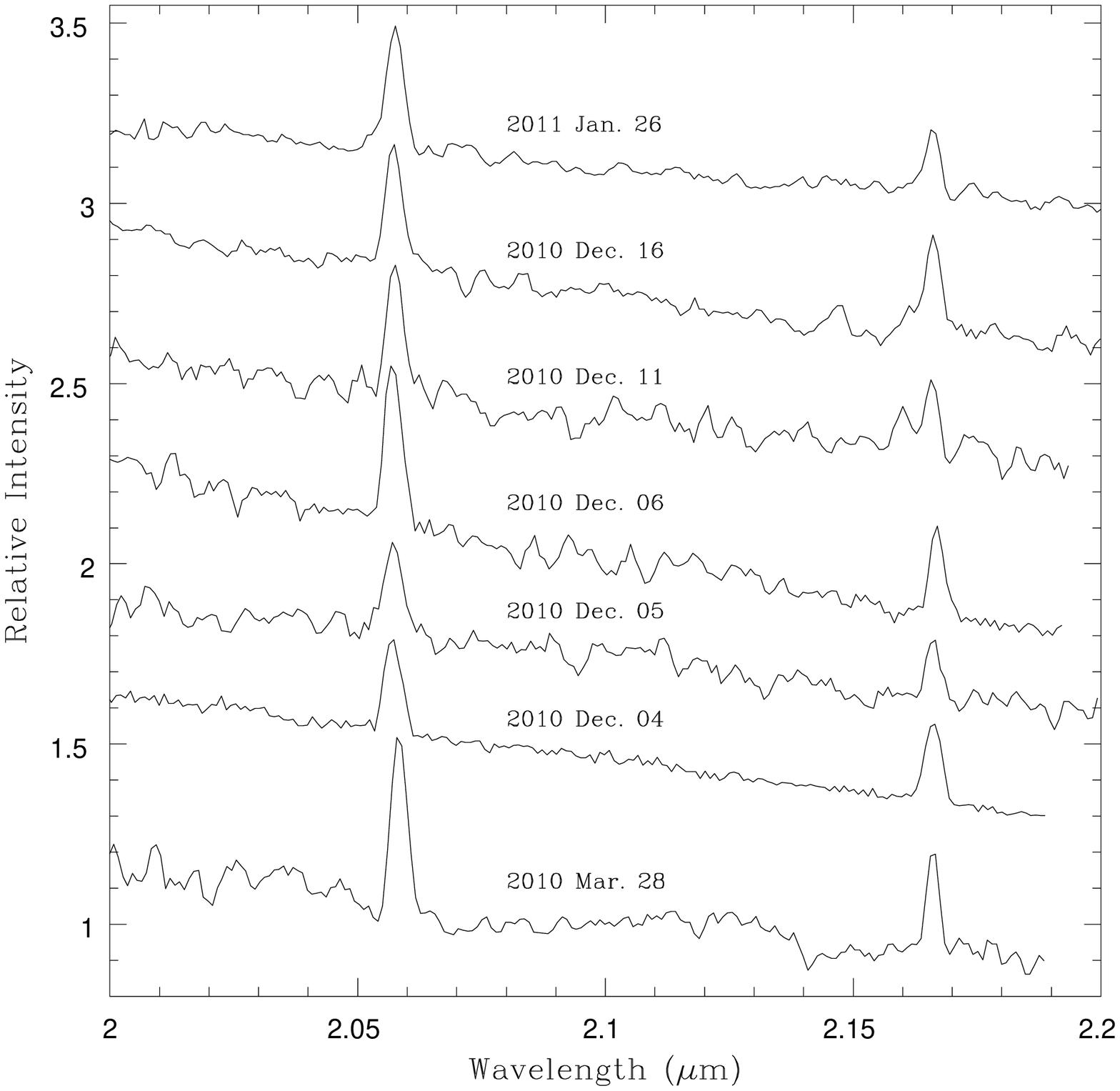}
\caption{The $K$-band spectra of A0535+262/HDE~245770 at different epochs with the
continuum being normalized to unity at 2.19 $\mu$m. }
\label{K-band}
\end{figure*}

\begin{sidewaystable}[h]
\caption{Equivalent widths of emission lines detected in the $JHK$ spectra of the Be star in A0535+262/HDE~245770 binary system.}
\begin{tabular}{@{}lccccccccccccccc@{}}
\hline
\hline
Date & Pa$\beta$ & Pa$\gamma$ & HeI & Br10 & HeI & Br11 & Br12 &Br13 & Br14 & Br15 & Br16 & Br17 & Br18 & Br$\gamma$ & HeI \\
\hline
2010 Mar. 28 	&-8.9	&-16.2	&---	&-7.3	&-4.5	&-7.1	&-6.9	&-6.1	&-7.2	&-4.7 &-7.2 &-5.6 &---  &-9.3  &-19.4\\
2010 Apr. 08	&-8.9	&-21.3	&-20.7	&-8.8	&-2.3	&-6.6	&-6.1	&-4.9	&-6.3	&-4.8 &-5.9 &-5.6 &-3.8 &---   &---  \\
2010 Oct. 31	&-3.9	&---	&---	&-5.7	&-1.8	&-6.9	&-2.3	&-3.7	&-3.9	&-2.7 &-2.8 &---  &-5.9 &---   &---  \\
2010 Dec. 04	&-12.2	&---	&---	&-8.9	&-1.3	&-8.3	&-4.3	&-9.5	&-5.2	&-4.7 &---  &---  &---  &-9.1  &-9.1 \\
2010 Dec. 05	&-4.4	&-17.0	&---    &-5.6	&-1.4	&-8.6	&-3.8	&-5.1	&-4.8	&-4.3 &-5.8 &-5.1 &-3.1 &-5.6  &-8.6 \\
2010 Dec. 06	&-10.5 	& --- 	&---   	&---	&-2.2 	&-8.5 	&-6.1 	&-4.8 	&-4.8 	&-5.4 &-5.2 &-5.2 &-3.4 &-8.2  &-13.7\\
2010 Dec. 11	&-9.3 	&---   	&---   	&-9.9 	&-2.8 	&-6.8 	&-5.9 	&-4.7 	&-6.7 	&-8.4 &-5.3 &-3.9 &---  &-7.7  &-12.8\\
2010 Dec. 16	&-4.2 	&---   	&---   	&-11.6 	&-2.9 	&-8.9 	&-7.4 	&-6.3 	&-5.4 	&-4.1 &-4.8 &-4.9 &-2.4 &-9.6  &-11.0\\
2011 Jan. 13	&-4.8 	&---   	&---   	&-5.4 	&-1.3 	&-5.7 	&-2.8 	&-2.8 	&-3.8 	&-3.3 &-6.4 &---  &---  &---   &---  \\
2011 Jan. 26	&-9.5 	&---   	&---   	&---	&-1.9 	&-4.3 	&-2.9 	&-4.8 	&-6.3 	&-3.2 &-5.4 &-5.8 &-2.4 &-6.9  &-13.3\\
2011 Feb. 09 	&-6.4 	&---   	&---   	&---  	&-1.6 	&-6.1 	&-5.3 	&-7.9 	&-6.4 	&-3.5 &-6.3 &-3.7 &-3.5 &-6.7  &-14.0\\
2011 Feb. 16 	&-6.8 	&---   	&---  	&-5.4 	&-1.8 	&-7.7 	&-5.4 	&-6.6 	&-6.8 	&-4.4 &-7.5 &-6.5 &-4.3 &-7.0  &-12.4\\
2011 Feb. 17	&-5.8 	&---   	&---  	&-7.4 	&-1.9 	&-7.8 	&-7.0 	&-5.4 	&-7.1 	&-5.2 &-5.8 &-4.5 &-3.1 &-7.1  &-9.9 \\
2011 Feb. 18	&-5.9 	&---   	&---  	&-5.9 	&-1.9 	&-9.4 	&-6.7 	&-5.8 	&-5.9 	&-4.5 &-5.8 &-4.8 &-4.1 &-7.0  &-10.5\\
2011 Feb. 19	&---  	&---   	&---  	&-6.5 	&-1.2 	&-7.3 	&-8.0 	&-4.4 	&-4.5 	&-4.5 &-7.7 &-4.8 &-3.0 &-6.2  &-11.2\\
2011 Feb. 27	&-4.8 	&---   	&---  	&-5.7 	&-1.2 	&-5.2 	&-6.4 	&-5.6 	&-7.3 	&-5.2 &-9.8 &---  &---  &-7.1  &-9.1 \\
2011 Feb. 28	&-6.5 	&---   	&---  	&-4.5 	&-2.7 	&-5.9 	&-5.9 	&-5.3 	&-6.1 	&-4.1 &-4.9 &-5.8 &-4.9 &-6.4  &-12.8\\
2011 Mar. 04	&-5.7 	&---   	&---  	&--- 	&-3.2 	&-8.3 	&-5.6 	&-5.0 	&-5.5 	&-5.9 &-8.8 &-4.2 &-3.5 &-8.9  &-12.0\\
2011 Mar. 05	&-7.9 	&-13.8 	&---  	&-8.5 	&-2.5 	&-8.8 	&-5.8 	&-4.8 	&-7.6 	&-4.3 &-6.4 &-4.3 &---  &-7.4  &-12.5\\
2011 Mar. 06	&-4.1 	&-21.0 	&---  	&-5.0 	&-3.3 	&-6.4 	&-4.7 	&-5.5 	&-7.6 	&-4.3 &-4.5 &-5.4 &---  &-9.3  &-11.8\\
2011 Mar. 10	&---  	&---   	&---   	&-7.1 	&-2.9	&-6.4 	&-5.3 	&-9.6 	&-7.1 	&-6.5 &-5.5 &-5.6 &-4.7 &-6.5  &-13.8\\
2011 Mar. 11	&-6.1 	&---   	&---   	&-6.6 	&-2.5	&-6.8 	&-5.9 	&-4.7 	&-7.2 	&-4.7 &-7.5 &-5.5 &-4.2 &---   &---  \\
2011 Mar. 14	&-6.1 	&-13.3 	&---   	&-8.7 	&-3.4	&-7.5 	&-3.5 	&-5.3 	&-5.9 	&-3.6 &-5.6 &-4.9 &-4.0 &-11.9 &-16.3\\
\hline
\hline
\end{tabular}
\\
Pa$\beta$=1.2818 ${\rm{\mu}}$m, Pa$\gamma$=1.0934 ${\rm{\mu}}$m, HeI=1.0830 ${\rm{\mu}}$m, Br10=1.7362 ${\rm{\mu}}$m, HeI=1.7002 ${\rm{\mu}}$m, Br11=1.6806 ${\rm{\mu}}$m,  Br12=1.6407 ${\rm{\mu}}$m, \\
Br13=1.6109 ${\rm{\mu}}$m, Br14=1.5881 ${\rm{\mu}}$m, Br15=1.5701 ${\rm{\mu}}$m, Br16=1.5545 ${\rm{\mu}}$m, Br17=1.5439 ${\rm{\mu}}$m, Br18=1.5342 ${\rm{\mu}}$m, Br$\gamma$=2.1655 ${\rm{\mu}}$m, HeI=2.0587 ${\rm{\mu}}$m
\end{sidewaystable}

\subsection{Emission lines in the $JHK$ spectra}

As mentioned earlier, the Be star in the A0535+262/HDE~245770 binary
system at $JHK$ magnitudes of $\sim$ 7.5--8.0 (Table 1) is a rather faint and
challenging target  for the 1.2 m telescope to get good quality near-IR
spectrum at the observed resolution of $\sim$1000. Therefore, our primary
aim was essentially to record the spectra to look for unusual or drastic
changes  rather than look for development of new lines or fine changes in
the line profiles. For example, we were interested to see whether striking 
phenomena like a complete disc-loss, that had earlier been reported for 
the object (Haigh et al. 1999), could be possibly effected by the neutron 
star's periastron passage. Such an effect could have been detected in our 
spectra. Further, we sought to see whether the observed line strengths 
changed majorly during the X-ray outburst indicating that line-emitting 
material of the disc had been majorly affected. Such large changes could 
again be expected to be detectable  in our spectra. Our spectra have thus 
been collected keeping these aims and restrictions in mind  and being aware 
that a deeper study of the near-IR spectra of A0535+262 typically need 
observations on much larger telescopes as shown by the study of Clark et 
al. (1998, 1999).

The $JHK$ spectra of the Be star in the A0535+262/HDE~245770 binary
system are presented in Figures~\ref{J-band}, ~\ref{H-band} \& ~\ref{K-band},
respectively. The left panels of Figures~\ref{J-band}, ~\ref{H-band} \&
~\ref{K-band} represent the $JHK$ spectra of the Be star since the onset of
the 2011 February--March giant X-ray outburst. The spectra in the right panels
of the figures, however, are during the rest of our observations that includes
the X-ray quiescent phase and a few epochs during minor X-ray outbursts (as
shown in Figure~\ref{x-lc}). The $JHK$ spectra of the Be star, throughout our
near-IR observation campaign, display expected emission lines of Paschen and
Brackett series lines from hydrogen. The details of the line identification
and corresponding equivalent widths are given in Table~2. The HeI line at
2.0581 $\mu$m~ is clearly detected in the $K$-band spectra during most of the
nights of our observation. The HeI line at 1.0830 $\mu$m, however, is 
detected in the $J$-band spectra during only a couple of nights of observation,
as extending the spectrum to cover this line requires a second positioning of 
the grating in the $J$-band which was not always possible. The HeI 1.7002 
$\mu$m line is also prominently seen in the $H$ band. The presence of HeI 
lines, Br$\gamma$ and other hydrogen lines in the $JHK$ spectra indicate 
that the Be star in the A0535+262/HDE~245770 binary system belongs to 
Group~I of the Clark \& Steele (2000) classification scheme and its 
spectral type is earlier than B3 which is consistent with its present 
classification of O9.7IIIe. A notable feature in the $H$-band spectra 
is the structure of the Br11 line at 1.6806 $\mu$m which is seen to be
distinctly different from  other Brackett series lines in terms of both 
width and shape. This is most likely caused due to the blending with the 
FeII 1.679 and FeII 1.687 $\mu$m lines which are known to be quite often 
present in the spectra of Be stars from the H band survey of Steele \& 
Clark (2001). 

Apart from these, changes in the shape of several emission 
line profiles are seen in the $JHK$ spectra of the Be/X-ray binary 
A0535+262/HDE~245770. At a few epochs during the X-ray outburst, the 
Pa$\beta$ and Br$\gamma$ lines appear to have structures in their profile 
and may also possibly be blended with other weak emission line. For example, 
Pa$\beta$ and Br$\gamma$ may be contaminated with HeI emission lines 
at 1.2748 and 2.1614 $\mu$m which are sometimes seen in the spectra of
other late O/B[e] stars (Hanson, Conti \& Rieke 1996; Clark et al. 1999).
However, it is difficult to be certain of line profile changes or the presence 
of such HeI lines in A0535+262 Be/X-ray binary system due to the 
low resolution and signal-to-noise 
ratio of the spectral data presented here. Significant variability in 
the emission line profiles has also been seen in the infrared spectra of 
Be/X-ray binary A0535+262/HDE~245770 (Clark et al. 1998b). Variability 
in line profiles was detected from the low and high resolution infrared
spectroscopy of the Be binary obtained over 1992--1995 that was interpreted 
as due to the changes in the circumstellar environment during this time. 
Strong similarity in the profiles of HeI 1.008, 2.058 $\mu$m, H$\alpha$ and 
Paschen series lines was also seen in the Echelle spectra of the Be star in 
this binary system. Optical high-dispersion spectroscopic monitoring 
observations of the Be binary system showed that the H$\alpha$ and H$\beta$ 
line profiles are variable over a period of 500 days (Moritani et al. 2011).

From our study, we find only marginal changes in the strengths of the
different lines (Table 2) during the binary orbital phases i.e. during
X-ray quiescent and X-ray outburst phases, of the Be/X-ray binary
A0535+262. The absence of any major changes in the values of the line
equivalent widths at entire orbital phases suggest that the line emitting
region in the circumstellar disc, that is closer to the Be star, is not
significantly affected by the periastron passage of the neutron star.

We also did a recombination Case B analysis of the HI lines using
predicted line strengths given in Storey and Hummer (1995).  Visual 
inspection of the $J$-band spectra straightaway shows that there is 
a considerable deviation from Case B conditions. As seen, the strength
of Pa$\gamma$ emission line at 1.0938 $\mu$m is larger or comparable 
to that of Pa$\beta$ at 1.2818 $\mu$m - the reverse of this behaviour 
is expected under case~B conditions. We also carried out the recombination 
case~B analysis for the HI Brackett series lines in the $H$ and $K$ band 
spectra of the Be companion for several epochs of observations at different 
orbital phases (during X-ray quiescent -- before periastron passage of the 
neutron star and X-ray outburst phase) of the A0535+262/HDE~245770 binary 
system. Although not shown here graphically there are considerable deviations 
from Case B conditions. For example, we specifically find that Br $\gamma$ which 
is expected to be considerably stronger than the higher  Br lines like Br10, 
11, 12 etc  is consistently observed to be weaker. These results indicate 
that the Br lines are optically thick causing thereby a deviation from the 
expected recombination case~B strengths. Optical depth effects are not 
unexpected given that  the electron densities in the circumstellar disc 
of Be stars is  generally high in the range of 10$^{10}$ cm$^{-3}$ -- 10$^{13}$
cm$^{-3}$ as reported by Steele \& Clark (2001).

\section{Summary}
Near-IR monitoring of the Be/X-ray binary system A0535+262/HDE~245770 during the
giant X-ray outburst in 2011 February--March and X-ray quiescent phases, shows a
$\sim$12\% reduction in the near-IR flux during the periastron passsage of the
neutron star. The increase in X-ray brightness during the X-ray outburst could
possibly be due to the evacuation of matter from the Be circumstellar disc which
was contributing ($\sim$12\% during the 2011 February--March X-ray outburst) to
the total near-IR emission from the Be binary system.  A series of $JHK$ 
spectra were taken during X-ray quiescent phase and the giant X-ray outburst 
in 2011 February--March to look for any major changes in the spectra -- no 
such changes were found to take place.

\section*{Acknowledgments}
The research work at Physical Research Laboratory is funded by the
Department of Space, Government of India. We thank Nafees Ahmad and Jinesh
Jain for help with some of the observations. This research has made use of
data obtained through HEASARC Online Service, provided by the NASA/GSFC, in
support of NASA High Energy Astrophysics Programs.

\end{document}